\documentclass[aps,prl,twoside,twocolumn,english,superscriptaddress,
	showkeys,showpacs,nofootinbib,tightenlines,floats,
	amssymb,amsmath,amsfonts]{revtex4}

\usepackage{babel}
\usepackage{amsmath}
\usepackage{amsfonts}
\usepackage{amssymb}
\usepackage{dsfont}

\begin{document}

%\begin{flushleft}
%IGC-09/2-2
%\end{flushleft}

\setcounter{page}{1}

\title{Peccei--Quinn mechanism in gravity and the nature of the Barbero--Immirzi parameter}

\author{\firstname{Simone} \surname{Mercuri}}
\email{mercuri@gravity.psu.edu}
\affiliation{Institute for Gravitation and the Cosmos, The Pennsylvania State University,\\ Physics Department, Whitmore Lab, University Park, PA 16802, USA}

\begin{abstract}
A general argument provides the motivation to consider the Barbero--Immirzi parameter as a field. The specific form of the geometrical effective action allows to relate the value of the Barbero--Immirzi parameter to other quantum ambiguities through the analog of the Peccei--Quinn mechanism.
\end{abstract}



\maketitle

\section{General Remarks}\label{GR-1}

One of the most successful attempts to construct a nonperturbative quantum theory of gravity is \emph{Loop Quantum Gravity} \cite{AshLew04,Rov04,Thi06}. Its classical starting point is the Ashtekar--Barbero canonical formulation of General Relativity (GR) \cite{Ash86-87-88,Bar95}, which, classically, corresponds to a modification of the Hilbert--Palatini (HP) action as demonstrated by Holst \cite{Hol96}. This modification consists in adding to the usual HP action a new term which vanishes on (half-)shell, i.e.\footnote{The signature throughout the paper is $+,-,-,-$; we set $c=\hbar=1$.}
\begin{align}\label{holst action}
\nonumber S\left[e,\omega\right]&=S_{\rm HP}\left[e,\omega\right]+S_{\rm Hol}\left[e,\omega\right]
\\
&=-\frac{1}{16\pi G}\int e_a\wedge e_b\wedge\left(\star R^{a b}+\beta R^{a b}\right)\,,
\end{align}
where $\beta$ is a constant known as Barbero--Immirzi (BI) parameter \cite{Imm97}. The parameter $\beta$ is not fixed by the theory and its origin is still debated \cite{Mer09p,DatKauSen09,Mer08,ChoTunYu05,GamObrPul99,RovThi98}. The quantum theory is, basically, the result of the Dirac quantization procedure \cite{Dir64} applied to the constraints of classical GR in the Ashtekar--Barbero formulation. Remarkably, by using holonomies and fluxes as fundamental variables in analogy with lattice quantum field theories, the quantum representation algebra can be rigorously formalized \cite{AshLew04,Rov04,Thi06} and a class of geometrical partial observables \cite{Rov04}, as the area and volume of a portion of spacetime, can be defined. Their associated operators, once suitably regularized, result to have a discrete spectrum depending on the BI parameter, $\beta$.

It is important to mention that the action (\ref{holst action}) is classically equivalent to the HP one, as the additional term vanishes as soon as the homogeneous second Cartan structure equation is solved. It goes without saying that this is no longer true if a source for torsion is present. The case of fermion fields coupled to gravity is particularly instructive and has been studied in many recent papers \cite{FreMinTak05,Ran05,PerRov06,Mer06,BojDas08}. Interestingly enough, it is easy to demonstrate that, in the presence of fermionic matter, the contribution coming from the Holst term can be compensated by a further modification in the fermionic sector \cite{Mer06}. Thus, the resulting effective theory is the usual Einstein--Cartan (EC) one plus a modification which can be associated with the so-called Nieh--Yan (NY) topological term \cite{NieYan82}.

This fact suggests a natural generalization of the Holst action (\ref{holst action}) to spacetimes with torsion, namely we can assume that the action describing the dynamics of the gravitational field is the following one \cite{Mer09p},
\begin{align}\label{new action for gravity}
\nonumber S_{\rm Grav}\left[e,\omega\right]=&\,S_{\rm HP}\left[e,\omega\right]+S_{\rm NY}\left[e,\omega\right]
\\
=&-\frac{1}{16\pi G}\int e_a\wedge e_b\wedge\star R^{ab}
\\\nonumber
&+\frac{\beta}{16\pi G}\int\left(T^a\wedge T_a-e_a\wedge e_b\wedge R^{a b}\right)\,,
\end{align}
where the second term is exactly the NY invariant. Remembering the definition of the torsion 2-form, $T^a=d e^a+\omega^a_{\ b}\wedge e^b$, the NY term can be rewritten as a total divergence, i.e.
\begin{equation}\label{Nieh--Yan}
\int\left(T^a\wedge T_a-e_a\wedge e_b\wedge R^{a b}\right)=\int d\left(e_a\wedge T^a\right)\,.
\end{equation}
Hence, in this new framework, the modification is a true topological term related to Pontryagin classes through the MacDowell--Mansouri connection \cite{ChaZan97}. The inclusion of matter is straightforward. As noted before, the case of spinor fields is particularly interesting: generating torsion in spacetime, they can, in fact, reveal the claimed properties of the NY term. 

According to the minimal prescription, the dynamics of the matter and gravity coupled system is described by the following action:
\begin{align}\label{action gravity-matter}
\nonumber S_{\rm Tot}=&\,S_{\rm HP}\left[e,\omega\right]+S_{\rm NY}\left[e,\omega\right]+S_{\rm D}\left[e,\omega,\psi,\overline{\psi}\right]
\\\nonumber 
=&-\frac{1}{16\pi G}\int e_a\wedge e_b\wedge\star R^{ab}
\\\nonumber
&+\frac{i}{2}\int\star e_a\wedge\left(\overline{\psi}\gamma^a D\psi-\overline{D\psi}\gamma^a\psi+\frac{i}{2}m e^a\overline{\psi}\psi\right)
\\
&+\frac{\beta}{16\pi G}\int\left(T^a\wedge T_a-e_a\wedge e_b\wedge R^{a b}\right)\,.
\end{align}
The covariant derivative $D$ contains the Lorentz valued connection $\omega^{a b}$, according to the following definition: $D\psi=d\psi-\frac{i}{4}\,\omega^{a b}\Sigma_{a b}\psi$;
$\Sigma^{a b}=\frac{i}{2}\left[\gamma^a,\gamma^b\right]$ being the generators of the Lorentz group. It is easy to demonstrate that pulling back the original action on the solution of the second Cartan structure equation one obtains the EC effective action, with its typical axial-axial four fermions interaction. In other words, the new modification does not affect the classical equations of motion, even though a source for torsion is present \cite{Mer06,DatKauSen09} (see also \cite{Kau08} for an analog result in supergravity theory); but, in general, it can affect the quantum theory introducing P and CP violating effects. It is worth noting that in this framework the BI parameter $\beta$ seems to play a role analogous to that played by the $\theta$-angle of QCD, as already argued in \cite{Mer08} (see also \cite{DatKauSen09,Mer09p}).

Recently, in a certain number of papers \cite{TavYun08,GomKra09,CalMer09,MerTav09,Fre91}, it has been proposed to promote the BI parameter to be a field rather than a constant. In particular, in \cite{CalMer09}, it was suggested that the interaction of the BI field with gravity was mediated by the NY term (\ref{Nieh--Yan}). As a result, in the effective theory, the BI field decouples from gravity and behaves exactly as a (pseudo) scalar field, whereas, in the previous models based on the Holst action \cite{TavYun08,GomKra09,Fre91}, it was $\phi=\sinh\beta$ to play the role of a scalar field rather than $\beta$ itself, complicating also the effective interaction with fermions \cite{GomKra09}. In this perspective, the NY modification appears more natural and, to a great extent, more appealing than the Holst one; though, a strong motivation to promote the BI parameter to a field is still missing.

In this letter, we suggest a possible motivation for considering the hypothesis that $\beta$ is actually a field. Specifically, in the next Section, it will be demonstrated that a well known divergent contribution to the chiral anomaly can be reabsorbed if $\beta$ is a new field of the theory. Moreover, by extracting the vacuum expectation value of the BI field,\footnote{Here vacuum has to be intended as pure gravitational field.} we argue that the analog of the Peccei--Quinn mechanism allows to relate the constant value of the BI parameter to other possible topological ambiguities, producing an interesting interaction between the physical BI field with gravity. The last Section is dedicated to a brief discussion of the possible physical consequences of this model.

\section{The BI field as a consequence of the chiral anomaly}

Let us consider the action (\ref{action gravity-matter}) where the fermionic sector describes ordinary spin-$1/2$ matter fields interacting with gravity. Specifically, let us assume that the fermionic action describes leptons as well as quarks. It is well known that the quark mass matrices $M$ resulting from the spontaneous breaking of the $SU(2)\times U(1)$ symmetry are neither diagonal nor Hermitian \cite{ItzZub,Wei96}. In order to diagonalize $M$ a chiral rotation is necessary. The parameter of the transformation depends on the number of flavors $n_f$ and on the ${\rm Arg}\det M$, specifically we have:
\begin{subequations}
\begin{align}
q_{R}\rightarrow q_R^{\prime}&=e^{\frac{i}{2n_f}{\rm Arg}\det M}q_R\,,
\\
q_{L}\rightarrow q_L^{\prime}&=e^{-\frac{i}{2n_f}{\rm Arg}\det M}q_L\,,
\end{align}
\end{subequations}
where $q_{R}$ and $q_{L}$ describes, respectively, a right and left handed quark of any flavor.
It turns out that such a chiral transformation introduces a divergent term in the effective theory.

In this respect, we recall that, in spacetime with torsion, the chiral rotation of the fermionic measure in the Euclidean path-integral generates, besides the usual Pontryagin class, a NY term, which diverges as the square of the regulator \cite{ChaZan97}, i.e.
\begin{align}\label{jacobian}\nonumber
\delta\psi\delta\overline{\psi}\rightarrow &\delta\psi\delta\overline{\psi}
\exp\left\{\frac{i}{8\pi^2}\int\alpha\left[R_{a b}\wedge R^{a b}\right.\right.
\\
&+2\Lambda^2\left.\left.\left(T_a\wedge T^a-e_a\wedge e_b\wedge R^{a b}\right)\right]\frac{}{}\!\right\}\,.
\end{align}
Above, $\Lambda$ denotes the regulator, while $\alpha$ is the parameter of the transformation.\footnote{The imaginary unit $i$ disappears in Minkowski space.} The robustness of this result has been confirmed by many independent calculations \cite{ObuMieBud97,Soo99,SooCha99}, performed by using different regularization procedures; however, it is worth mentioning that it has been claimed in \cite{KreMie01} that, in fact, the divergent term is irrelevant for the chiral anomaly. In this respect, we stress that the main claim in \cite{KreMie01} was based on an untenable observation, mainly motivated by a classical ``on-shell'' condition, which, as noted in \cite{ChaZan01}, cannot be extended to quantum regimes and to general torsional spacetimes. So, according to Eq.(\ref{jacobian}), the resulting effective action, 
\begin{align}\label{Simo1}\nonumber
S_{\rm Tot}&\left[e,\omega,\psi,\overline{\psi}\right]=S_{\rm HP}\left[e,\omega\right]+S_{\rm D}\left[e,\omega,\psi,\overline{\psi}\right]
\\\nonumber
+&\frac{1}{16\pi G}\left(\beta+\frac{4G}{\pi}\,\alpha\Lambda^2\right)\int\left(T^a\wedge T_a-e_a\wedge e_b\wedge R^{a b}\right)
\\
+&\frac{1}{8\pi^2}\,\alpha\int R_{a b}\wedge R^{a b}\,,
\end{align}
($\alpha=\frac{1}{n_f}{\rm Arg}\det M$) diverges as soon as we try to remove the regulator. 

It is important to note that the divergence can be reabsorbed by promoting the $\beta$ parameter to be a field and then renormalizing it. According to this observation, let us introduce in the fundamental action the field $\beta(x)$, which interacts with gravity through the NY term, i.e.
\begin{align}\label{fundamental action}
\nonumber S_{\rm Tot}&\left[e,\omega,\psi,\overline{\psi},\beta\right]=S_{\rm HP}\left[e,\omega\right]+S_{\rm D}\left[e,\omega,\psi,\overline{\psi}\right]
\\
&+\frac{1}{16\pi G}\int\beta(x)\left(T^a\wedge T_a-e_a\wedge e_b\wedge R^{a b}\right)\,.
\end{align}
By assuming that (\ref{fundamental action}) is the fundamental action for gravity, we can immediately eliminate the possible divergence resulting from the diagonalization of the quarks mass matrix by renormalizing the field $\beta(x)$. In fact, by using the invariance of the new action under a shift of $\beta(x)$, without affecting the dynamical content of the theory, we can incorporate the divergence in the definition of a new field $\beta^{\prime}\left(x\right)=\beta\left(x\right)-\frac{4G}{\pi}\,\alpha\Lambda^2$. 

In this new framework the BI parameter can be naturally associated to the expectation value of the field $\beta(x)$, namely $\beta_0=\left<\beta(x)\right>$.

Dynamically, the new action (\ref{fundamental action}) is no longer equivalent to the EC one, in fact the presence of the new field $\beta(x)$ modifies the expression of the torsion 2-form, which reflects on the torsionless effective theory. In this respect, by varying action (\ref{fundamental action}) with respect to the Lorentz valued 1-form $\omega^{a b}$ and manipulating it, we obtain the following structure equation,
\begin{equation}\label{torsion tensor}\nonumber
d e^a+\omega^{a}_{\ b}\wedge e^b=T^a=\epsilon^{a}_{\ b c d}\left(\frac{1}{2}\eta^{b f}\partial_{f}\beta-2\pi G J^b_{(A)}\right)e^c\wedge e^d\,,
\end{equation}
where $J_{(A)}=e_a J^a_{(A)}=\sum_{f=1}^{n}e_a\overline{\psi}_f\gamma^a\gamma^5\psi_f$ is the fermionic axial current (the sum is extended to all the spin-$1/2$ matter fields contained in the action and described by the collective symbols $\psi$ and $\overline{\psi}$). The connection $\omega^{a b}$ can be easily calculated. We have $\omega^{a b}={}^{\circ}\!{\omega}^{ab}(e)+K^{ab}(e,\beta,\psi,\overline{\psi})$,
where ${}^{\circ}\!{\omega}^{ab}(e)$ is the usual Ricci spin connection, while the contortion 1-form $K^{ab}$ is explicitly given by the following expression:
\begin{equation}
K^{a b}(e,\beta,\psi,\overline{\psi})=\epsilon^{a b}_{\ \ c d}e^c\left(2\pi G J^d_{(A)}-\frac{1}{2}\eta^{d f}\partial_{f}\beta\right)\,.
\end{equation}
It is worth noting that for $\beta={\rm const.}$ the solution reduces to the ordinary solution of the EC theory. This fact suggests that, if $\beta={\rm const.}$, the NY modification does not affect the classical effective action, even though a source for torsion, e.g. spinor fields, is considered, exactly as claimed above and here demonstrated as a particular case in a more general framework. More generally, when the parameter $\beta$ is promoted to be a field, the NY term in the action ceases to be topological generating a modification with respect to the EC theory. Interestingly enough, in order to preserve the usual transformation properties of the torsion tensor under the action of the Lorentz group and, consequently, its geometrical interpretation, the $\beta(x)$ field has to be a pseudo-scalar, as can be easily understood by considering expression (\ref{torsion tensor}). In other words, the pseudo-scalar nature of $\beta(x)$ is not assumed \emph{a priori}, but is a geometrical consequence of the theory.

Let us now explicitly extract the effective action. By calculating the torsion contributions contained in the different pieces of the total action, we have:
%\begin{widetext}
\begin{subequations}
\begin{align}
\nonumber &S_{\rm HP}\left[e,\omega\right]=S_{\rm HP}\left[e,{}^{\circ}\!{\omega}\right]-\frac{3}{2}\,\pi G\int dV \eta_{a b}J^a_{(A)}J^b_{(A)}
\\
&+\frac{3}{4}\int\star J_{(A)}\wedge d\beta-\frac{3}{32\pi G}\int \star d\beta \wedge d\beta\,,
\\
\nonumber &S_{\rm D}\left[e,\omega,\psi,\overline{\psi}\right]=S_{\rm D}\left[e,{}^{\circ}\!{\omega},q_f,\overline{q}_f\right]
\\
&+3\pi G\int dV \eta_{a b}J^a_{(A)}J^b_{(A)}
-\frac{3}{4}\int\star J_{(A)}\wedge d\beta\,,
\\
\nonumber &S_{\rm NY}\left[e,\omega\right]=\frac{1}{16\pi G}\int\beta d\left(e_a\wedge T^a\right)
\\
&=-\frac{3}{4}\int\star J_{(A)}\wedge d\beta+\frac{3}{16\pi G}\int\star d\beta\wedge d\beta\,,
\end{align}
\end{subequations}
%\end{widetext}
where torsionless geometrical objects are distinguished by a circle.\footnote{For the sake of notational clarity, we note that $\star d\beta=\star e^a\partial_a\beta=\frac{1}{3!}\epsilon^{a}_{\ b c d}e^b\wedge e^c\wedge e^d\partial_a\beta$. The same is valid for the 1-form $J_{(A)}$.} 

As discussed above, in general, the effective action contains P and CP violating terms, included in the effective action via the parameter $\tilde{\theta}=\theta+\frac{1}{8\pi^2}\,\alpha$, i.e.
\begin{align}\label{STE}
\nonumber &S_{\rm eff}=S_{\rm HP}\left[e,\underset{\circ}{\omega}\right]+S_{\rm D}\left[e,\underset{\circ}{\omega},\psi,\overline{\psi}\right]
\\\nonumber
&\ +\frac{3}{2}\,\pi G\int d V \eta_{a b}J^a_{(A)}J^b_{(A)}+\frac{1}{2}\int\star d\tilde{\beta}(x)\wedge d\tilde{\beta}(x)
\\
&\ +\frac{\tilde{\theta}}{8\pi^2}\int R^{ab}\wedge R_{ab}-\sqrt{3\pi G}\int\star J_{(A)}\wedge d\tilde{\beta}(x)\,,
\end{align}
where to be as general as possible also a non-vanishing vacuum angle, $\theta$, has been considered. Remarkably the divergent contribution in (\ref{Simo1}) no longer affects the effective action because reabsorbed through the shift symmetry in the field $\beta$. For convenience, we have defined the new field $\tilde{\beta}(x)=\frac{1}{4}\sqrt{\frac{3}{\pi G}}\beta(x)$, so that it has the dimension of an energy and the kinetic term in the action has the usual numerical factor. The net effect of torsion in the effective action (\ref{STE}) is the presence of the four fermions interaction as well as the appearance of the kinetic and interacting terms for the new dynamical field $\tilde{\beta}(x)$. In other words, the dynamics of the field $\tilde{\beta}(x)$ is completely determined by the geometrical content of the theory; in particular, torsion generates an interesting physical interaction between the pseudo-scalar field $\tilde{\beta}(x)$ and the fermionic axial current $J_{(A)}$, described by the last term of (\ref{STE}). Let us now focus the attention exactly on this interaction term. As stressed before, the axial current fails to be conserved because of the non-invariance of the path-integral measure under chiral rotations.\footnote{We recall that the axial current fails to be conserved even in the classical theory, the deviation being proportional to the mass of the fermionic particles. Here we consider only the quantum contribution to the chiral anomaly.} Specifically, considering only gravity, we now have that $d\star J_{(A)}=-\frac{1}{8\pi^2}R_{ab}\wedge R^{ab}$. This allows us to contain the two terms in the last line of the effective action (\ref{STE}) in a single expression, namely 
\begin{equation}\label{STE2}
S_{\rm CP}\left[e,\beta\right]=\frac{1}{8\pi^2}\int\left(\tilde{\theta}+\sqrt{3\pi G}\,\tilde{\beta}(x)\right)R^{ab}\wedge R_{ab}\,.
\end{equation}
The term above is the only CP violating contribution to the effective action. Furthermore, it represents a non-trivial potential for the field $\tilde{\beta}$, with a stationary point in $\tilde{\theta}+\sqrt{3\pi G}\,\tilde{\beta}(x)=0$ preserving $P$ and $CP$ symmetries. By using the same argument valid for the axion (see, e.g., \cite{Wei96}), we can eliminate the term (\ref{STE2}) from the effective action, by a chiral transformation affecting the fermionic mass term. Assuming for simplicity that the fermionic contribution to the effective action contains only the up and down quarks, we can easily derive an effective action for the pion and the BI field. For the details we refer the reader to the standard Literature \cite{Wei96} and to \cite{MerTav09}, describing here only the main results. From the quadratic part in $\tilde{\beta}$ of the effective Lagrangian, we can extract the mass of the BI field, namely $m_{\beta}\approx 10^{-12}{\rm eV}$. The higher orders in the effective Lagrangian represent a potential for the massive field $\tilde{\beta}$. The presence of a non-trivial potential for the BI field fixes its expectation value, i.e. $\beta_0=-4/3\tilde{\theta}$, then the value of the BI parameter can be directly related to the $\tilde{\theta}$ ambiguity of the theory, which in general depends on the non-trivial global structure of the local gauge group. Moreover, we note that the physical field $\tilde{\beta}_{\rm phys}=\tilde{\beta}-<\!\tilde{\beta}\!>$ interacts with gravity through the Pontryagin density, so that the effective gravitational theory resembles a Chern--Simons modified gravity \cite{JacPi.03}, largely studied in the Literature (see \cite{KonMatTan07,AleYun07-1,AleYun07-2} and references therein).

The tiny mass of the BI field $\tilde{\beta}$ and the possible existence of the axion induce to consider a more realistic physical situation, in which the other interactions are taken into account as well as the necessity to solve the so-called \emph{strong $CP$ problem}. These considerations are contained in the next Section.

\section{Discussion}\label{Discussion}

In order to get some more realistic and interesting physical predictions from this model, we cannot neglect the presence of another pseudo-scalar particle $a(x)$, which, through the Peccei--Quinn mechanism \cite{PecQui77,Pec98}, naturally solves the strong $CP$ problem (for gravitational analogues see \cite{DesDufIsh80,Ale05}). In other words, to delineate a more complete picture, we have to consider in the effective action also the presence of an additional pseudo-scalar degree of freedom, $a(x)$, and take into account that, through the chiral anomaly, both $\tilde{\beta}$ and $a(x)$ interact with the gauge fields.

A general argument suggests that the fields $a(x)$ and $\tilde\beta(x)$ naturally interact via the chiral anomaly with the gauge bosons in a linear combination, schematically \cite{PosRitSko08}:
\begin{align}\nonumber
\mathcal{L}_{\rm int}&=\left(\frac{\beta}{2g_{\beta}}+\frac{a}{2g_{a}}\right){\rm tr}G\wedge G+\left(\frac{\beta}{2f_{\beta}}+\frac{a}{2f_{a}}\right)F\wedge F
\\
&+\left(\frac{\beta}{2r_{\beta}}+\frac{a}{2r_{a}}\right)R^{a b}\wedge R_{a b}\,,
\end{align}
where $G$ is the curvature 2-form of the $SU(3)$ valued connection 1-form associated to the strong interaction, while $F$ is the electromagnetic field strength. The mechanism that solves the strong $CP$ problem and simultaneously generates an anomaly-induced mass is peculiar and only one linear combination of the two pseudo-scalar fields acquires a mass \cite{AnsUra82}. At an effective level, this fact implies that besides the usual QCD term for the massive physical axion (the mass depending on the energy scale of the interaction), one has a massless additional pseudo-scalar field, $\Phi$, which interacts with the electromagnetic as well as the gravitational field as follows:
\begin{align}\label{Sara1}
\mathcal{L}_{\rm int}&=\mathcal{L}_{\rm axion}+\frac{\Phi}{2f_{\Phi}}F\wedge F
+\frac{\Phi}{2r_{\Phi}}R^{a b}\wedge R_{a b}\,,
\end{align}
where $f^{-1}_{\Phi}$ and $r^{-1}_{\Phi}$ denote the scale of the respective interactions.

The above interaction (\ref{Sara1}) has interesting physical consequences. The presence of the coupling with photons, in fact, induces a rotation of the polarization angle, $\varepsilon$, of an electromagnetic wave \cite{HarSik92}; particularly interesting in this direction could be the study of the polarization anisotropies of CMB as recently claimed in \cite{PosRitSko08}. Moreover, the same mechanism induces a rotation of the polarization angle of the gravitational waves through the interaction with the gravitational term, as demonstrated in the context of Chern--Simons modified gravity \cite{AleFinYun07,AleYun08} (see also \cite{AleMar05,AleGatJam06,AlePesShe07,Ale08} for applications to cosmology). Interestingly enough, recently a black hole solution for Chern--Simons modified gravity has been found \cite{PreYun09}, representing a natural scenario where this model can be tested.

\vspace{.2cm}
{\bf Acknowledgments.}
This research was supported in part by NSF grant PHY0854743, The George A. and Margaret M. Downsbrough Endowment and the Eberly research funds of Penn State. The author thanks A. Ashtekar, G. Calcagni, A. Randono and V. Taveras for discussions and is grateful to S. Amato for having indirectly suggested the presented idea.

\end{document}